\documentclass[a4paper,11pt]{article}
\pdfoutput=1 

\usepackage{jheppub} 
\usepackage{bbm}

\newcommand\cm{\mathcal{M}}

\newcommand\beal{\begin{align}}
\newcommand\nn{\nonumber}
\newcommand\bbone{\ensuremath{\mathbbm{1}}}

\newcommand{\eq}[1]{\begin{equation}#1\end{equation}}
\newcommand{\spl}[1]{\begin{split}#1\end{split}}

\newcommand{\mcal}{\mathcal{M}}

\newcommand{\ncal}{\mathcal{N}}

\newcommand{\G}{\Gamma}
\newcommand{\g}{\gamma}
\newcommand{\e}{\epsilon}
\newcommand{\we}{\widetilde{\eta}}

\newcommand{\p}{\partial}
\renewcommand{\O}{\Omega}

\renewcommand{\a}{\alpha}
\renewcommand{\b}{\beta}
\renewcommand{\o}{\omega}
\renewcommand{\t}{\theta}

\newcommand{\boxedeq}[1]{
\begin{equation}
\fbox{
\rule[0.7cm]{0pt}{0pt}
$#1$
\rule[-0.45cm]{0pt}{0pt}
}
\end{equation}
}

\def\d{\text{d}}

\def\slashchar#1{\setbox0=\hbox{$#1$}           
\dimen0=\wd0                                 
\setbox1=\hbox{/} \dimen1=\wd1               
\ifdim\dimen0>\dimen1                        
\rlap{\hbox to \dimen0{\hfil/\hfil}}      
#1                                        
\else                                        
\rlap{\hbox to \dimen1{\hfil$#1$\hfil}}   
/                                         
\fi}

\def\Re           {{\rm Re\hskip0.1em}}

\title{IIB supergravity on manifolds with $SU(4)$ structure\\
and generalized geometry}

\author{Dani\"{e}l Prins and}
\author{Dimitrios Tsimpis}

\affiliation{Universit\'{e} de Lyon\\
UMR 5822, CNRS/IN2P3, Institut de Physique Nucl\'{e}aire de Lyon\\
4 rue Enrico Fermi,
F-69622 Villeurbanne Cedex,  France\\}
\emailAdd{dlaprins@ipnl.in2p3.fr}
\emailAdd{tsimpis@ipnl.in2p3.fr}

\abstract{We consider $\ncal=(2,0)$ backgrounds of  IIB supergravity on eight-manifolds $\mcal_8$ with strict $SU(4)$ structure. We give the explicit solution to the Killing spinor equations as a set of algebraic relations between irreducible $su(4)$ modules of the fluxes and the torsion classes of $\mcal_8$. One consequence of supersymmetry is that $\mcal_8$ must be complex.
We show that the conjecture of \href{http://arxiv.org/abs/1010.5789}{\tt arxiv:1010.5789} concerning the correspondence between background supersymmetry equations in terms of generalized pure spinors and generalized calibrations for admissible static, magnetic D-branes, does not capture the full set of supersymmetry equations.
We identify the missing constraints and express them in the form of a single pure-spinor equation which is well defined for generic $SU(4)\times SU(4)$ backgrounds. This additional equation is given in terms of a certain  analytic continuation of the generalized calibration form for codimension-2 static, magnetic D-branes.}

\begin{document}
\maketitle
\flushbottom

\setcounter{footnote}{0}
\renewcommand{\thefootnote}{\arabic{footnote}}
\setcounter{section}{0}


\section{Introduction and summary}\label{introduction}\label{sec1}

In searching for string theory vacua it is often useful
to consider the problem within the low-energy approximation of supergravity.
On the other hand generic solutions of supergravity involve non-zero flux, the presence of which has necessitated the use of new mathematical tools. In particular the framework of generalized geometry \cite{hitch,gual} has proven to be very well suited for the description of supergravity backgrounds in the presence of flux. It has lead to important insights into the general structure of flux vacua and underlies much of the recent progress in the construction of effective actions, sigma models, as well as supersymmetry breaking and non-geometry. For a review of generalized geometry  for physicists see  \cite{koer}.

For type II supergravity backgrounds of the form $\mathbb{R}^{1,3}\times\cm_6$ in particular, the conditions for an $\ncal=1$ (two complex supercharges) supersymmetric bosonic background can be expressed as a set of first-order differential equations for two complex pure spinors of $Cl(6,6)$ \cite{gran}.
Furthermore an important connection was noted in \cite{calmart}: the aforementioned pure-spinor equations are in one-to-one correspondence with the differential conditions obeyed by the (generalized) calibration forms of all admissible supersymmetric\footnote{A D-brane is called supersymmetric if it doesn't break the supersymmetry of the background. In the present paper all D-branes we consider are assumed to be supersymmetric.} static, magnetic D-branes in that background. One may hope to promote this correspondence to an organizing
principle for flux vacua. To that end one would like to know to what extent
this correspondence may be applicable to more
general setups.

In \cite{patalong} the one-to-one correspondence between supersymmetry equations (in pure-spinor form) and calibration forms for static, magnetic D-branes was shown to also hold for  $\ncal=1$ backgrounds of the form $\mathbb{R}^{1,5}\times\cm_4$ (four complex supercharges). Based on these results the authors of \cite{patalong} conjectured that the correspondence should also hold for certain $\ncal=(2,0)$ type II backgrounds of the form $\mathbb{R}^{1,1}\times\cm_8$ (one complex supercharge)\footnote{More generally the conjecture of \cite{patalong} was formulated
for type II supersymmetric backgrounds of the form  $\mathbb{R}^{1,d-1}\times\cm_{10-d}$, with $d=2,4,6,8$, preserving $2^{d/2-1}$ complex supercharges.}. In particular it was noted in \cite{patalong} that in a type II background of the form $\mathbb{R}^{1,1}\times\cm_8$ the only admissible  static, magnetic D-branes are spacetime-filling -- where spacetime is identified with the `external' $\mathbb{R}^{1,1}$ part. The supersymmetry equations were then conjectured to be given by:
\boxedeq{\spl{\label{conjp}
  \d_H \left( e^{2 A- \Phi } \text{Re} \Psi_1 \right) &= e^{2A} \star_8 \sigma(F)\\
   \d_H \left( e^{2 A- \Phi} \Psi_2 \right) &= 0
~,}}
where $d_H\equiv d+H\wedge$, $\Psi_{1,2}$ are generalized pure spinors of $Cl(8,8)$, and the generalized calibration for spacetime-filling static, magnetic D-branes is given by a linear combination of  $e^{2 A- \Phi } \text{Re} \Psi_1$ and $e^{2 A- \Phi} \Psi_2$.

It is worth emphasizing
that the conjecture of \cite{patalong} concerns supersymmetric backgrounds for which the `internal' part (along $\mcal_8$) of the Killing spinors is given in terms of {\it pure} Weyl spinors (ordinary, not generalized) of $Cl(8)$. Given that Weyl spinors of $Cl(8)$ are not necessarily pure, the supersymmetric backgrounds considered in \cite{patalong} are not the most general. In the present paper we consider
$\ncal=(2,0)$ backgrounds of the form $\mathbb{R}^{1,1}\times\cm_8$ such that
the Killing spinor ansatz a) is of the form considered in \cite{patalong} and b) is `strict' (also known as `rigid') $SU(4)$. These assumptions imply the existence of a nowhere-vanishing pure spinor on $\mcal_8$
and thus, as reviewed in appendix \ref{sec3}, the reduction of the structure group of $\mcal_8$ to $SU(4)$.\footnote{It should be emphasized that supersymmetry does not necessarily imply the reduction of the structure group of $\mcal_8$ to a subgroup of SO(8): it is rather the structure group of an auxiliary nine-manifold $\mcal_8\times S^1$ which reduces in general  \cite{tsimpism}. In the context of M-theory this point has been further developed in \cite{lazb}; see \cite{mcorist} for an explicit example.}

The reduction of the structure group is equivalent to the existence on $\mcal_8$
of a real nowhere-vanishing two-form $J$ and a complex nowhere-vanishing four-form $\Omega$ obeying certain compatibility conditions. It follows in particular that the covariant spinor derivative can be expressed in terms of $SU(4)$ torsion classes, where the latter parameterize the failure of closure of $(J,\Omega)$ and thus the departure from the Calabi-Yau condition.
Decomposing all fluxes in terms of irreducible modules of $su(4)$ we are then able to re-express the Killing spinor equations in terms of a set of algebraic relations between the different components of the fluxes and the torsion classes. The solution to the Killing spinor equations is given in
\eqref{solution} below. An immediate consequence of supersymmetry is that $\mcal_8$ must be complex, as can be seen by the vanishing of the first two torsion classes. Moreover it can be seen that the solution (\ref{solution}) to the strict $SU(4)$-structure ansatz includes the class of compactifications on conformally Calabi-Yau fourfolds as well as backgrounds for which $\mcal_8$ is conformally K\"{a}hler.

The procedure described in the previous paragraph goes under the name of $G$-structures, see e.g.~\cite{joyce}; in the context of supergravity it was initiated in \cite{ma,mc} and has proven very fruitful in the search
for explicit solutions. The case of IIB supersymmetric backgrounds has been considered in all generality in \cite{0507087}, using the ``spinorial geometry'' approach developped in \cite{0410155}. These techniques have proven particularly powerful in classifying backgrounds with maximal supersymmetry \cite{0505074, 0604079}. In the special case considered here where there is a reduction of the structure group to $G=SU(4)$, and in order to make contact with generalized geometry and calibrations, it is more suitable to reformulate the problem in the language of $SU(4)$ structures.

Having explicitly solved the supersymmetry equations we were then able to test the conjecture of \cite{patalong} given in \eqref{conjp} above, for the special case of a strict $SU(4)$-structure ansatz. We have found that the conjecture of \cite{patalong} captures only part of the background supersymmetry equations. We have identified the `missing' equations and have shown that they can be succintly put in the form a single generalized pure-spinor equation:
\boxedeq{\label{missing}
  \d^{\mathcal{I}_2}_H \left( e^{- \Phi } \text{Im} \Psi_1 \right) = F
~,}
where $\mathcal{I}_2$ is the generalized almost complex structure associated with the pure spinor $\Psi_2$ (the relation between generalized almost complex structures and generalized pure spinors is reviewed in section \ref{app:gengeom}). As follows from the second line of \eqref{conjp},  $\mathcal{I}_2$ is in fact integrable, and the operator $\d^{\mathcal{I}_2}$ appearing in the equation above can be written in terms of the generalized Dolbeault operator associated with $\mathcal{I}_2$, $\d^{\mathcal{I}_2}_H\equiv
i(\bar{\partial}^{\mathcal{I}_2}_H-\partial^{\mathcal{I}_2}_H)$.

As we have seen, it follows from the analysis of the supersymmetry equations for the case of strict $SU(4)$ structure considered here that the internal manifold $\mcal_8$ is complex. Moreover the completely holomorphic and antiholomorphic parts of $H$ vanish, so that $H=H^{(2,1)}+H^{(1,2)}$ with respect to the complex structure.
The generalized Dolbeault operator can then be seen to reduce to $\partial_H\equiv\partial+H^{(2,1)}\wedge$ and thus the operator $\d^{\mathcal{I}_2}_H$ appearing in \eqref{missing} is simply  $i(\bar{\partial}_H-\partial_H)$, where $\bar{\partial}_H=\bar{\partial}+H^{(1,2)}\wedge$ is the complex conjugate of $\partial_H$. Expressed in terms of $\d^{\mathcal{I}_2}_H$, \eqref{missing} remains well-defined beyond the subclass of strict $SU(4)$ structure for which it was derived in the present paper. It is then plausible, correcting \cite{patalong}, to formulate a new conjecture stating that for $\ncal=(2,0)$ type IIA/B backgrounds of the form $\mathbb{R}^{1,1}\times\cm_8$, with spinor ansatz given in (\ref{introspindecompa}), (\ref{normeta}) below, the background supersymmetry equations are equivalent to \eqref{conjp}, \eqref{missing}.

As already mentioned, the polyform $e^{- \Phi }\text{Im} \Psi_1$  in (\ref{missing}) cannot be associated with static, magnetic D-branes.\footnote{Euclidean (instantonic) D-branes cannot be supersymmetric in Minkowskian spacetimes since they correspond to time-dependent configurations without null Killing vectors. Supersymmetric configurations which would correspond to instantonic D-branes may be constructed in Euclidean supergravity, see e.g. \cite{green}. In the present setup they would be codimension-2 with respect to $\mathbb{R}^{1,1}$.}
 Interestingly however  it was noted in \cite{patalong} that for $d=4,6,8$ the polyform $e^{(d-2)A- \Phi }\text{Im} \Psi_1$ does appear and is associated with  static, magnetic D-branes which wrap some cycle in $\mcal_8$ and are codimension-2 with respect to the external part $\mathbb{R}^{1,d-1}$ of the ten-dimensional background (cf. table 1 of \cite{patalong}). Of course in the present case $d=2$ and D-branes which are codimension-2 in $\mathbb{R}^{1,1}$  cannot be static, since by definition static D-branes must wrap the time direction. Nevertheless  $e^{- \Phi }\text{Im} \Psi_1$  can be considered as the `analytic continuation'
to $d=2$ of the generalized calibration form  $e^{(d-2)A- \Phi }\text{Im} \Psi_1$ for codimension-2 static, magnetic D-branes.

The plan of the remainder of this paper is as follows: The $\mathbb{R}^{1,1}\times\mcal_8$, type IIB $\mathcal{N} =(2,0)$  backgrounds we consider and the associated Killing spinor ansatz are described in detail in section \ref{sec2}. The solution of the Killing spinor equations in terms of $SU(4)$ structures is given in \eqref{solution} in section \ref{secsusy}. The equivalence of the latter to the generalized pure-spinor equations \eqref{conjp}, \eqref{missing} is explained in section \ref{seccal}.

To improve the presentation of the results, we have moved most technical details to the appendices. Appendix \ref{app1} includes our spinor and gamma-matrix conventions. Appendix \ref{sec3} sets up the framework of eight-manifolds with $SU(4)$ structures and works out the decomposition of fluxes in terms of $su(4)$ irreducible modules. The expression of the spinor covariant derivative in terms of $SU(4)$ torsion classes is given in \eqref{toreta}; to our knowledge this is the first time it appears explicitly in the literature.

Additional material on generalized geometry and calibrations is included in section \ref{app:calibrations}. In particular the equivalence of the generalized twisted Dolbeault operator $\partial_H^{\mathcal{I}_2}$ to the ordinary twisted Dolbeault operator $\partial_H$ for the case of strict $SU(4)$ backgrounds is explained in section \ref{app:gengeom}.

\section{$\mathcal{N}=(2,0)$ supersymmetric backgrounds}\label{sec2}

The general case of IIB supersymmetric backgrounds has been analyzed  in \cite{0507087} in the framework of spinorial geometry \cite{0410155}. For our purposes it will useful to reformulate the problem in the language of $SU(4)$ structures. We will consider ten-dimensional type IIB backgrounds which
are topologically  direct products
of the form $\mathbb{R}^{1,1}\times\cm_{8}$. The manifold $\cm_{8}$  is assumed to be Riemannian and spin.
The ten-dimensional metric reads:
\eq{\label{10dmetric}
\d s^2 = e^{2A}\d s^2(\mathbb{R}^{1,1})+\d s^2(\cm_{8})
~,}
where the warp factor $A$ is taken to only depend on
the coordinates of the internal manifold $\cm_{8}$.
We will also assume that not all
RR charges are zero; the case with zero RR charges has already
been analyzed in \cite{gauntlett}. The most general
RR charges respecting the two-dimensional Poincar\'e symmetry of our setup are of the form:\footnote{We follow the `democratic' supergravity conventions of \cite{march}, see appendix A therein, except for the ten dimensional Hodge-star operator $*_{10}$ which we define as
\eq{\spl{\label{star10}
*_{10}\omega_{p}\, =\, \frac{1}{p!(10-p)!}\sqrt{-g}\,\epsilon_{M_1\ldots M_{10}}\omega^{M_{11-p}\ldots M_{10}} \d x^{M_1}\wedge \ldots\wedge \d x^{M_{10-p}}\nonumber~,
}}
with $\epsilon_{01\ldots 9}=1$. }
\eq{\label{fluxan}
F^{\mathrm{tot}}=\mathrm{vol}_2\wedge F^{\mathrm{el}}+F
~,}
where $\mathrm{vol}_2$ is the unwarped volume element of $\mathbb{R}^{1,1}$,
 and we are using polyform notation.
We denote by $F$ the `magnetic' RR charges with indices
along the internal space $\cm_{8}$.
The ten-dimensional Hodge duality relates $F$ to the `electric' RR charges
via:
\eq{
F^{\mathrm{el}}=e^{2A}\star_{8}\sigma(F)
~,}
where the Hodge star above is with respect to the internal metric, and
the involution $\sigma$ acts by inverting the order of the form indices.
%
%

Following \cite{patalong} we consider $\ncal=(2,0)$ backgrounds where the Killing spinors of the ten-dimensional background are given by:
\eq{\label{introspindecompa}
\epsilon_i=\zeta\otimes\eta_i+\zeta^c\otimes\eta_i^c
~,}
with $i=1,2$, so that $\epsilon_{1,2}$ are Majorana-Weyl spinors of $Spin(1,9)$ of the same chirality;
$\zeta$ is a complexified\footnote{\label{complexified}We use the term `complexified' for a Weyl spinor with complex components. The term `complex Weyl spinor'
is reserved for Weyl spinors whose complex conjugate has opposite chirality.}, positive-chirality Killing spinor of $\mathbb{R}^{1,1}$. It corresponds to one complex supercharge -- two positive-chirality Majorana-Weyl spinors of $Spin(1,1)$, each of which corresponds to one real supercharge -- hence the spinor ansatz above is indeed $\ncal=(2,0)$ in two Minkowski dimensions; the precise form of the complex conjugate spinors $\zeta^c$, $\eta^c$ on the right hand side of the above equation as well as our spinor conventions are explained in appendix \ref{app1}.
As in \cite{patalong}, $\eta_{1,2}$ are pure spinors of $\mcal_8$  of equal norm:
\eq{\label{normeta}
|\eta_1|^2=|\eta_2|^2=\mathrm{const}\times e^{\frac{1}{2}A}~,
}
where $A$ is the warp factor of $\mathbb{R}^{1,1}$ as appears in \eqref{10dmetric}. This condition can be seen to follow from the requirement that the
background admits kappa-symmetric branes which do not break the
background supersymmetry, see e.g. \cite{koermart}.

In this paper we will require in addition a strict $SU(4)$ ansatz. This means that the two pure spinors are proportional to each other: $\eta_1\propto \eta_2$, so that we may set $\eta_2=e^{i\t}\eta_1$, where $\t$ is a real function on $\mcal_8$. It follows that we may choose the following parameterization:
\eq{\label{spindecompeta}
\eta_1=\alpha\eta~,~~~\eta_2=\alpha e^{i\t}\eta~,
}
where $\eta$ is a pure, positive-chirality Weyl spinor of $Spin(8)$ of unit norm and $\alpha$, $\t$ are real functions on $\mcal_8$.

Note that the Killing spinor ansatz
(\ref{introspindecompa}) considered here does not allow AdS$_2$ for the external part
of the metric. This can readily be seen from the Killing spinor equation for AdS$_2$: $\nabla_\mu \zeta_+ = W \gamma_\mu \zeta_-$, where $\zeta_+$, $\zeta_-$ are positive-, negative-chirality spinors respectively and $W$ is proportional to the inverse radius of AdS$_2$. On the other hand the irreducible chiral representation of $Spin(1,1)$ is real, which implies that the spinors $\zeta$, $\zeta^c$ that appear in (\ref{introspindecompa})
have the same (positive) chirality. Therefore the only way the Killing spinor equation for AdS$_2$ can be satisfied is in the limit $W\rightarrow 0$, which corresponds to flat Minkowski space.

\section{Supersymmetry in terms of $SU(4)$ structures}\label{secsusy}

We now proceed to solving the Killing spinor equations for the bosonic IIB backgrounds described in section \ref{sec2}, using the machinery of $G$-structures. In our conventions the Killing spinor equations are given by:
\eq{\spl{\label{kse}
\delta \lambda^1 &= \left(\underline{\p}\phi + \frac{1}{2} \underline{H} \right) \epsilon_1+ \left( \frac{1}{16} e^{\phi} \G^M \underline{F}^{\mathrm{tot}}\G_M \G_{11} \right) \epsilon_2 = 0 \\
\delta \lambda^2 &= \left( \underline{\p }\phi - \frac{1}{2} \underline{H} \right) \epsilon_2 - \left( \frac{1}{16} e^{\phi} \G^M \sigma(\underline{F}^{\mathrm{tot}})\G_M \G_{11} \right) \epsilon_1 = 0 \\
\delta \psi^1_M &= \left( \nabla_M + \frac{1}{4} \underline{H}_M \right) \epsilon_1 + \left( \frac{1}{16} e^{\phi} \underline{F}^{\mathrm{tot}}\G_M \G_{11} \right) \epsilon_2 = 0 \\
\delta \psi^2_M &= \left( \nabla_M - \frac{1}{4} \underline{H}_M \right) \epsilon_2 - \left( \frac{1}{16} e^{\phi} \sigma(\underline{F}^{\mathrm{tot}})\G_M \G_{11} \right) \epsilon_1 = 0 \;,
}}
where for any $(p+q)$-form $S$ we define:
\eq{
\underline{S}_{M_1\dots M_q}\equiv\frac{1}{p!}\Gamma^{N_1\dots N_p}
{S}_{N_1\dots N_pM_1\dots M_q}
~.}
We then decompose all fluxes into $su(4)$ modules using \eqref{fluxan} and the formul\ae{} of section \ref{sec4}. We also decompose all gamma matrices as in Appendix \ref{app1} and we use the ten-dimensional Killing spinor ansatz (\ref{introspindecompa}), (\ref{spindecompeta}).
Finally, using \eqref{toreta} and taking into account that $\zeta$ is a Killing spinor of $\mathbb{R}^{1,1}$, so that $\nabla_{\mu}\zeta=0$, the Killing spinor equations reduce to the following set of algebraic relations:
\boxedeq{\spl{\label{solution}
W_1 &= W_2 = 0 \\
W_3 &= i e^\phi (\cos\t f^{(2,1)}_3  - i \sin\t f^{(2,1)}_5) \\
W_{4} &= \frac{2}{3}\p^+  (\phi- A )\\
W_{5} &=  \p^+ (\phi - 2 A +  i \t)\\
\a &= e^{ \frac{1}{2} A}\\
\tilde{f}^{(1,0)}_{3} &=  \tilde{f}^{(1,0)}_{5} = \tilde{h}^{(1,0)}_{3} =0 \\
h_{1}^{(1,0)} &= 0\\
h_{3}^{(1,0)} &= \frac{2}{3} \p^+ \t \\
f_{1}^{(1,0)} &=-i \p^+( e^{- \phi}\sin\t )\\
f_{3}^{(1,0)} &= - \frac{i}{3}e^{2A} \p^+( e^{-2A- \phi}\cos\t )\\
f_{5}^{(1,0)} &=\frac{1}{3}e^{-4A} \p^+( e^{4A- \phi}\sin\t )\\
f_{7}^{(1,0)} &=e^{-2A} \p^+( e^{2A- \phi}\cos\t )\\
h^{(2,1)} &= e^\phi (- \cos\t f^{(2,1)}_5  + i \sin\t f^{(2,1)}_3)
~.
}}
The solution above is parameterized by the real scalar fields $\t$, $A$, $\phi$, and the $(2,1)$-forms $f_{3}^{(2,1)}$, $f_{5}^{(2,1)}$; we have also absorbed a real constant in the definition of $\a$. We use the notation $S^{(p,q)}$ for a form which is of $(p,q)$-type with respect to the almost complex structure of $\mcal_8$, while for any scalar $T$, $\partial^+T$, $\partial^-T$ denote the projections of $\d T$ to its (1,0), (0,1) parts respectively.
As explained in appendix \ref{sectorsion}, it immediately follows from the vanishing of the torsion classes $W_1$, $W_2$ that the almost complex structure of $\mathcal{M}_8$ is in fact integrable, and thus $\mcal_8$ is a complex manifold. We may then introduce complex coordinates and identify $\partial^+T$ with $\partial T$ and $\partial^-T$ with $\bar{\partial}T$.

From the definition of the torsion classes in (\ref{torj}) it can be seen that under a Weyl transformation of the metric, $g\rightarrow e^{2\chi}g$, the torsion classes transform as follows:
\eq{\label{conformal}
W_2\rightarrow e^{2\chi}W_2~,~~~
W_3\rightarrow e^{2\chi}W_3~,~~~
W_4\rightarrow W_4+2\partial^+\chi~,~~~
W_5\rightarrow W_5+4\partial^+\chi
~,}
while $W_1$ is invariant. The manifold $\mcal_8$ is K\"{a}hler if and only if all torsion classes are zero with the  possible exception of $W_5$. From (\ref{conformal}) it thus follows that the condition for a conformally K\"{a}hler space is that $\mathrm{Re}W_4$ is exact and $W_i=0$ for $i=1,2,3$. Similarly, the condition for a Calabi-Yau space is that all torsion classes vanish; the condition for conformally Calabi-Yau is thus that  $\mathrm{Re}W_4$ is exact, $2W_4-W_5=0$ and $W_i=0$ for $i=1,2,3$. By inspection of (\ref{solution}) we see that the conformally K\"{a}hler as well as the conformally Calabi-Yau condition are easily satisfied for non-trivial fluxes.

\section{Supersymmetry in terms of generalized calibrations}\label{seccal}

In this section we present the details of the proof of the equivalence of the Killing spinor equations for the $\mathcal{N}=(2,0)$ strict-$SU(4)$ background described in section \ref{secsusy} and the set of generalized pure-spinor equations
(\ref{conjp}), (\ref{missing}). Some further explanatory material on generalized geometry and calibrations can be found in appendix \ref{app:calibrations}.

As already mentioned, for the backgrounds we are considering here the structure group of the tangent bundle of $\mcal_8$ is reduced to $SU(4)$. From the point of view of generalized geometry this is then a special case of generalized $SU(4)\times SU(4)$ structures described in terms of a pair of compatible,  nowhere-vanishing pure spinors.
The compatible pair of pure spinors is constructed explicitly in terms of the internal spinors of $\mcal_8$ as follows, cf. \eqref{purespinors}:
\eq{\spl{\label{4.1}
  \Psi_1 &=- e^{- i \t} e^{- i J}\\
  \Psi_2 &=- e^{i \t} \Omega ~,
}}
with $\eta_{1,2}$ as in (\ref{spindecompeta}), where the Clifford map was used  together with the definitions \eqref{b3} of $J$, $\Omega$ as spinor bilinears.

The calibration equations \eqref{conjp} proposed in \cite{patalong} were conjectured to be equivalent to the supersymmetry equations, in analogy with the known results in $d=4$ \cite{calmart} and $d=6$ \cite{patalong} external dimensions. Let us first note that in $d=2$ the NSNS three-form $H$ can have an external part that does not break two-dimensional Poincar\'{e} invariance, as in \eqref{b19}. This would have no clear interpretation from the  point of view of generalized geometry. Fortunately both the supersymmetry equations (\ref{solution}) and the pure-spinor equations (\ref{conjp}) imply $h_1=0$.
Indeed, (\ref{conjp}) are equivalent to the following set of equations:
\eq{\spl{\label{conjp1}
W_1 &= W_2 = 0 \\
W_3 &= i e^\phi \left(\cos\t f^{(2,1)}_3  - i \sin\t f^{(2,1)}_5\right) \\
W_{5} &=   \p^+ ( \phi-2A +  i \t)\\
\tilde{f}^{(1,0)}_{3} &=  \tilde{f}^{(1,0)}_{5}  = h^{(1,0)}_1= \tilde{h}^{(1,0)}_{3} =0 \\
f_{1}^{(1,0)} &= ie^{- \phi} \left( \cos\t (\p^+ \t - 3 h^{(1,0)}_{3} ) +  \sin\t( 2 \p^+ A -\p^+ \phi + 3 W_{4})\right)\\
f_{3}^{(1,0)} &= - i e^{- \phi} \left( \cos\t (\p^+ \phi - 2 \p^+ A  - 2 W_{4} ) + \sin\t (\p^+ \t - 2  h^{(1,0)}_{3} )\right)\\
f_{5}^{(1,0)} &= e^{- \phi} \left( \cos\t (\p^+ \t - h_{3}^{(1,0)} ) - \sin\t(\p^+ \phi-2 \p^+ A  - W_{4}) \right)\\
f_{7}^{(1,0)} &= e^{- \phi} \left( \cos\t (2 \p^+A  - \p^+ \phi) - \sin\t( \p^+ \t) \right)\\
h^{(2,1)} &= e^\phi \left(- \cos\t f^{(2,1)}_5  + i \sin\t f^{(2,1)}_3\right)
\;.
}}
It can readily be verified that (\ref{conjp1}) is consistent with the supersymmetry solution (\ref{solution}) but is missing a number of constraints.
As it turns out, these are equivalent to imposing:
\eq{\label{missingdc}
i(\bar{\partial}_H-\partial_H) \left( e^{- \Phi } \text{Im} \Psi_1 \right) = F
~,
}
in addition to (\ref{conjp1}), where
$\partial_H\equiv\partial+H^{(2,1)}\wedge$ is the ordinary twisted Dolbeault  operator and $\bar{\partial}_H=\bar{\partial}+H^{(1,2)}\wedge$ is its complex conjugate.
Indeed, it can be seen that (\ref{missingdc}) implies
in addition the following set of equations:
\eq{\spl{\label{missing1}
f_{1}^{(1,0)} &= i e^{- \phi} \left(\sin\t \p^+ \phi -  \cos\t (\p^+ \t)\right)\\
f_{3}^{(1,0)} &= i e^{- \phi} \left( \cos\t (\p^+ \phi - W_{4} )
+ \sin\t (\p^+ \t - h^{(1,0)}_{3} )\right)\\
f_{5}^{(1,0)} &= e^{- \phi}
\left(\sin\t( \p^+ \phi - 2 W_{4}) -\cos\t (\p^+ \t - 2
h_{3}^{(1,0)} ) \right)\\
f_{7}^{(1,0)} &= e^{- \phi} \left( \cos\t (\p^+ \phi - 3 W_{4})
+  \sin\t( \p^+ \t - 3  h_{3}^{(1,0)}) \right)~.
}}
Finally, the supersymmetry equations (\ref{solution}) can be seen to be equivalent to (\ref{conjp1}), (\ref{missing1}), provided one imposes
in addition the norm condition $\a = e^{\frac{1}{2} A}$.

As explained in appendix \ref{app:gengeom}, the operator on the
left hand side of (\ref{missingdc}) can be replaced by $\d^{\mathcal{I}_2}_H$, where
\eq{\label{x}\d^{\mathcal{I}_2}_H\equiv
i(\bar{\partial}^{\mathcal{I}_2}_H-\partial^{\mathcal{I}_2}_H)~
}
is given in terms of the generalized twisted Dolbeault operator defined in (\ref{dolbeault}), or alternatively as in (\ref{c12}).
Although this may seem as overkill, expressing (\ref{missingdc}) in terms of $\d^{\mathcal{I}_2}_H$ puts this equation in the form of \eqref{missing} which is well-defined for any $SU(4)\times SU(4)$-structure background (in which case $\mcal_8$ is not necessarily a complex manifold) and not only for strict $SU(4)$ backgrounds.

\section{Conclusions}\label{secconcl}

We have considered a subclass of the $\ncal=(2,0)$ supersymmetric $\mathbb{R}^{1,1}\times\mcal_8$ backgrounds of \cite{patalong}: that of IIB flux backgrounds with strict $SU(4)$ structure. We have recast the supersymmetry equations in terms of a set of algebraic relations between $su(4)$ irreducible modules of the fluxes and the torsion classes. We have seen that the class of conformally Calabi-Yau fourfolds, as well as that of conformally K\"{a}hler eight-dimensional manifolds are particular solutions to the supersymmetry equations.

In the present paper we did not consider the full set of supergravity equations of motion, but have rather restricted our analysis to the supersymmetry equations. It is known that under certain conditions, integrability theorems guarantee that imposing supersymmetry together with the (generalized) Bianchi identities for the forms implies the equations of motion of all NSNS fields \cite{lt,0507087,gaunteqs}. These integrability theorems have been extended for supersymmetric backgrounds that include calibrated branes \cite{kt}, as well as to non-supersymmetric backgrounds \cite{march}.
Expressed in the form of \eqref{solution} the solution to the Killing spinor equations, together with the integrability theorems, should facilitate the search for new explicit type IIB flux vacua. For the case of e.g. group manifolds and cosets with invariant $SU(4)$-structures, the framework of the present paper may be better suited than those of spinorial geometry \cite{0410155,0507087}. It would be interesting to pursue this further.

Having explicitly solved the supersymmetry equations we were able to test the conjecture of \cite{patalong} concerning the correspondence between background supersymmetry equations in terms of generalized pure spinors and generalized calibrations for admissible static, magnetic D-branes. We have found that the conjecture of \cite{patalong} misses a number of constraints; we have shown that these are equivalent to a single pure-spinor equation, given in (\ref{missing}).

Although strictly-speaking not necessary for the strict $SU(4)$ structure backgrounds considered here, \eqref{missing} has been expressed in terms of the twisted generalized Dolbeault operator. In this form it is well-defined for
generic $SU(4)\times SU(4)$ backgrounds.
It is then natural to replace the conjecture of \cite{patalong} by the statement that eqs.~(\ref{conjp}), (\ref{missing}) are exactly equivalent to the Killing spinor equations for $\ncal=(2,0)$ supersymmetric $\mathbb{R}^{1,1}\times\mcal_8$ backgrounds of type IIA/B supergravity, where the Killing spinor ansatz is given by (\ref{introspindecompa}) with $\eta_{1,2}$ equal-norm pure spinors on $\mcal_8$ obeying (\ref{normeta}). It would be interesting to test this conjecture beyond the case of strict $SU(4)$ IIB backgrounds considered here. One possible avenue may be to try to exploit the results of \cite{Tomasiello:2011eb,1110.0627} on type II backgrounds and generalized geometry.

Where do our results leave the background supersymmetry/generalized calibrations correspondence? As we have seen, if our new conjecture is verified, this correspondence can be rescued in a modified form: eqs.~(\ref{conjp}) are in one-to-one correspondence with
static, magnetic D-branes in the background, while (\ref{missing}) corresponds to the analytic continuation to two external spacetime dimensions (in the sense discussed in section \ref{introduction}) of the generalized calibration form for codimension-2 static, magnetic D-branes.


\appendix

\section{Spinor and gamma matrix conventions}\label{app1}

For a spinor $\psi$ in any dimension we define:
\eq{\widetilde{\psi}\equiv\psi^{Tr}C^{-1}~,}
where $C$ is the charge conjugation matrix. In Lorentzian signatures, we also define
\eq{\overline{\psi}\equiv\psi^{\dagger}\G^{0}~.}
In all dimensions the Gamma matrices are taken to obey
\eq{
(\G^M)^{\dagger}=\G^0\G^M\G^0~,
}
where the Minkowski metric is mostly plus. Antisymmetric products of
Gamma matrices are defined by
\eq{
\G^{(n)}_{M_1\dots M_n}\equiv\G_{[M_1}\dots\G_{M_n]}~.
}

\subsection*{Two Lorentzian dimensions}

The charge conjugation matrix in $1+1$ dimensions satisfies
\eq{
C^{Tr}=-C; ~~~~~~ (C\g^\mu)^{Tr}=C\g^\mu; ~~~~~~ C^*=-C^{-1}~.
}
The fundamental (one-dimensional, chiral) spinor representation is real.
In this paper we work with a complexified chiral spinor $\zeta$ (i.e. one complex degree of freedom). We define:
\eq{
\zeta^c\equiv\g_0C\zeta^*~.
}
The chirality matrix is defined by
\eq{
\g_3\equiv{{-}}\g_0\g_1 ~.
}
The Hodge-dual of an antisymmetric product of
gamma matrices is given by
\eq{
\star\g_{(n)}\g_3={{- (-1)^{\frac{1}{2}n(n+1)}}}\g_{(2-n)}~.
\label{hodge2}
}

\subsection*{Eight Euclidean dimensions}

The charge conjugation matrix in $8$ dimensions satisfies
\eq{
C^{Tr}=C; ~~~~~~ (C\g^\mu)^{Tr}=C\g^\mu; ~~~~~~ C^*=C^{-1}~.
}
The fundamental (eight-dimensional, chiral) spinor representation is real.
In this paper we work with a complexified chiral spinor $\eta$ (i.e. eight complex degrees of freedom). We define:
\eq{
\eta^c\equiv C\eta^*~.
}
The chirality matrix is defined by
\eq{
\g_9\equiv\g_1\dots\g_8 ~.
}
The Hodge-dual of an antisymmetric product of
gamma matrices is given by
\eq{
\star\g_{(n)}\g_9=(-)^{\frac{1}{2}n(n+1)}\g_{(8-n)}~.
\label{hodge8}
}

\subsection*{Ten Lorentzian dimensions}

The charge conjugation matrix in $1+9$ dimensions satisfies
\eq{
C^{Tr}=-C; ~~~~~~ (C\G^M)^{Tr}=C\G^M; ~~~~~~ C^{*}=-C^{-1}~.
}
The fundamental (16-dimensional, chiral) spinor representation
$\e$ is real, where we define the reality condition by
\eq{
\overline{\e}=\widetilde{\e}~.
}
The chirality matrix is defined by
\eq{
\G_{11}\equiv{{-}}\G_0\dots\G_9 ~.
}
We decompose the ten-dimensional Gamma matrices as
\eq{
\left\{ \begin{array}{ll}
\G^{\mu}=\g^\mu\otimes \bbone &, ~~~~~\mu=0, 1\nn\\
\G^m=\g_3\otimes \g^{m-1} &, ~~~~~ m=2\dots 9
\end{array} \right.
~.}
It follows that
\eq{
C_{10}=C_2\otimes C_8;~~~~~~ \G_{11}=\g_3\otimes\g_9~.
}
The Hodge-dual of an antisymmetric product of
gamma matrices is given by
\eq{
\star \G_{(n)} \G_{11}= {{- (-1)^{\frac{1}{2}n(n+1)}}}      \G_{(10-n)}~.
}

\section{$SU(4)$ structures}\label{sec3}

As we will now review a nowhere-vanishing complex, chiral, pure spinor $\eta$ of unit norm in eight euclidean dimensions defines an $SU(4)$ structure. In eight euclidean dimensions not every complex chiral spinor is pure: the property of purity is equivalent to the condition
\eq{\label{p1}
\we\eta=0~.}
Let $\eta_R$, $\eta_I$ be the real, imaginary part of $\eta$ respectively.
We will impose the normalization:
\eq{\label{b2}
\eta=\frac{1}{\sqrt{2}}(\eta_R+i\eta_I)~;~~~~~\we_R\eta_R=\we_I\eta_I=1~,}
so that $\tilde{\eta}^c\eta=1$, and (\ref{p1}) is equivalent to $\eta_R$, $\eta_I$ being orthogonal to each other: $\we_R\eta_I=\we_I\eta_R=0$.

Let us  define a real two-form $J$ and a complex self-dual four-form $\Omega$ through the spinor bilinears
\eq{\spl{\label{b3}
iJ_{mn}&=\widetilde{\eta^c}\g_{mn}\eta\\
\Omega_{mnpq}&=\we\g_{mnpq}\eta~.
}}
It can then be shown by Fierzing that these forms obey:
\eq{\spl{
J\wedge\Omega&=0\\
\frac{1}{16}\Omega\wedge\Omega^*&=\frac{1}{4!}J^4 =\mathrm{vol}_8~,
}}
up to a choice of orientation,
and hence define an $SU(4)$ structure.
The reduction of the structure group  can alternatively be seen as from the fact that $Spin(6)\cong SU(4)$ is the stabilizer inside $Spin(8)$ of the pair of orthogonal Majorana-Weyl unit spinors $\eta_R$, $\eta_I$.

Raising one index of $J$ with the metric defines an almost complex structure:
\eq{
J_m{}^pJ_p{}^n=-\delta_m^n
~.}
Using the almost complex structure
we can define the projectors
\eq{
(\Pi^{\pm})_m{}^n\equiv\frac{1}{2}(\delta_{m}{}^{n}\mp i J_m{}^n)
~,}
with respect to which $\Omega$ is holomorphic
\eq{
(\Pi^{+})_m{}^i\Omega_{inpq}=\Omega_{mnpq}~; ~~~~~(\Pi^{-})_m{}^i\Omega_{inpq}=0 ~.
}
Further useful relations are given in appendix \ref{app2}.

In eight dimensions the Clifford algebra $Cl(8)$ is equivalent to the set $\mathbb{R}[16]$ of real $16\times 16$ matrices. With complex coefficients the
gamma matrices generate $Cl(8)\otimes\mathbb{C}\cong\mathbb{C}[16]$. Since a complex Dirac spinor in eight dimensions can be thought of as a vector of $\mathbb{C}^{16}$ and $\mathbb{C}[16]$ acts transitively on $\mathbb{C}^{16}-\{0\}$, any complex Dirac spinor can be expressed as an element of $Cl(8)$, with complex coefficients, acting on the non-vanishing spinor $\eta$.

More explicitly, let $\xi_+$, $\xi_-$ be arbitrary positive-, negative-chirality complexified spinors (cf. footnote \ref{complexified}) respectively. Using equations (\ref{fierzsu}), it follows from the previous paragraph that $\xi_\pm$ can be expressed as:
\eq{\spl{
\xi_+&=\varphi~\!\eta+\chi~\!\eta^c+\varphi_{mn}\g^{mn}\eta^c\\
\xi_-&={{\lambda_{m}}} \g^{m}\eta+\chi_{m}\g^{m}\eta^c
~,
\label{xiexp}}}
where $\varphi$, $\chi$ are complex scalars, $\varphi_m$, $\chi_m$ are complex
(0,1)-, (1,0)-forms respectively and $\varphi_{mn}$ is a complex (2,0)-form. As a consistency check, we note that the arbitrary positive-chirality spinor $\xi_+$ is parametrized by eight complex degrees of freedom: two complex d.o.f.s from the complex scalars $\varphi$, $\chi$ plus six complex d.o.f.s from the complex (2,0)-form  $\varphi_{mn}$. Similarly $\xi_-$ is parametrized by four plus four complex d.o.f.s coming from the (0,1)-, (1,0)-forms $\varphi_m$, $\chi_m$ respectively.

\subsection{Torsion classes}\label{sectorsion}

The intrinsic torsion $\tau$ (see e.g. \cite{gauntlett} for a review) transforms in the $\Lambda^1(\mcal_8)\otimes su(4)^\perp$, where $su(4)^\perp$ is the complement of the adjoint of $su(4)$ inside the adjoint of $so(8)$. It follows that
\eq{\spl{
\tau&\in(\bf{4}\oplus\bf{\bar{4}})\otimes(\bf{1}\oplus \bf{6}\oplus\bf{6})\nn\\
&\sim(\bf{4}\oplus\bf{\bar{4}})\oplus(\bf{20}\oplus\bf{\bar{20}})
\oplus(\bf{20}\oplus\bf{\bar{20}})
\oplus(\bf{4}\oplus\bf{\bar{4}})\oplus(\bf{4}\oplus\bf{\bar{4}})
~,}}
in a decomposition in terms of irreps of $su(4)$. We then decompose $\tau$ in five `torsion classes' $W_1,\dots, W_5$, according to the second line on the right-hand side above. These torsion classes are the obstructions to the closure of the forms $J$, $\Omega$. Explicitly we will choose the following
parameterization:\footnote{We define the contraction
between a $p$-form $\varphi$ and a $q$-form $\chi$, $p\leq q$, by
\eq{
\varphi\lrcorner\chi=\frac{1}{p!(q-p)!}\varphi^{m_1\dots m_p}
\chi_{m_1\dots m_p n_1\dots n_{q-p}}\d x^{n_1}\wedge\dots\wedge\d x^{n_{q-p}}
~.}
Once the normalization of the $W_1$ term on the right-hand side of the first equation in (\ref{torj}) is fixed, the $W_1$ term on the right-hand side of the second equation can be determined as follows: Starting from $\d(J\wedge\Omega)=0$ we substitute for $\d J$, $\d\Omega$ using (\ref{torj}), taking (\ref{jvol}), (\ref{omvol}) into account and noting that
$W_2\wedge J\wedge J=0$ since $W_2$ is primitive.}
\eq{\spl{\label{torj}
\d J&=W_1\lrcorner\Omega^*+W_3+W_4\wedge J+\mathrm{c.c.}\\
\d\Omega&=\frac{8i}{3}W_1\wedge J\wedge J+W_2\wedge J+W_{5*}\wedge\Omega
~,}}
where $W_1$,  $W_4$, $W_5\sim\bf{4}$ are complex (1,0)-forms and $W_2$, $W_3\sim\bf{20}$ are complex traceless (2,1)-forms.

Equivalently, the torsion classes are the obstructions to the spinor $\eta$ being covariantly constant with respect to the Levi-Civita connection. Explicitly we have:
\boxedeq{\spl{\label{toreta}
\nabla_m \eta
&= \left( \frac{3}{4}W_{4m} - \frac{1}{2} W_{5m} - \mathrm{c.c.}\right) \eta +\frac{i}{24} \O^*_{mnkl}  W_{1}^{n} \g^{kl}\eta\\
& +\left(
-\frac{i}{16} W_{2mkl}
- \frac{1}{32}  \O_{mnkl} W_{4}^{n*}
+ \frac{i}{64} W^*_{3mnp} \O^{np}{}_{kl}
 \right) \g^{kl}\eta^c~.
}}
This can be seen as follows. From the discussion around (\ref{xiexp}) and the fact that $\nabla_m\eta$ transforms in the $\bf{8}\otimes\bf{8^+}$ of $so(8)$, we can expand
\eq{\label{toretaprov}
\nabla_m\eta=\varphi_m \eta + {{\vartheta_m}} \eta^c + \Psi_{m,pq} \Omega^{pqrs} \gamma_{rs} \eta^c~,}
for some complex coefficients $\varphi_m$, $\vartheta_m\sim{\bf 4}\oplus{\bf \bar{4}}$, $\Psi_{m,pq}\sim({\bf 4}\oplus{\bf \bar{4}})\otimes{\bf 6}$. Furthermore we decompose:
\begin{align}
\Psi_{m,pq} &= \Omega^*_{mpqr}A^r +  \Omega_{pq}^{*\phantom{pq}rs} \tilde{\varphi}_{rsm}+ (\Pi^+)_{m[p}B^*_{q]} + (\Pi^+)_{m}^{\phantom{m}n} \psi^*_{npq}
\end{align}
where $A,B \sim{\bf 4}$ are complex (1,0)-forms and $\tilde{\varphi},\psi \sim{\bf 20}$ are complex traceless (2,1)-forms.
Multiplying (\ref{toretaprov}) on the left with $\widetilde{\eta^c}\g_{ij}$ and $\we\g_{ijk}$, antisymmetrizing in all indices  in order to form $\d J$ and $\d \Omega$ respectively as spinor bilinears and  comparing with (\ref{torj}) then leads to \eqref{toreta}.

As can be seen from \eqref{torj}, the obstruction to having an integrable  almost complex structure is given by $W_1$, $W_2$. Conversely, if  $W_1$, $W_2$ vanish one can use \eqref{toreta} to show that the Nijenhuis tensor vanishes and thus the almost complex structure is integrable.

\subsection{Tensor decomposition}\label{sec4}

Under an $so(8)\rightarrow su(4)$ decomposition the one-, three-form of $so(8)$ decompose respectively as:
\eq{\spl{
\bf{ 8}&\rightarrow \bf{ (4\oplus \bar{4})}\nn\\
\bf{ 56}&\rightarrow \bf{(4\oplus \bar{4})\oplus (4\oplus 20)\oplus (\bar{4}\oplus \bar{20}) }~.
}}
Explicitly we decompose the RR tensors as follows.

{$\bullet$  Real one-form}
\eq{\label{dec1}F_m=f^{(1,0)}_{1|m}+\mathrm{c.c.}~,}
where $f^{(1,0)}_{1|m}\sim \bf{4}$ is a complex (1,0)-form with respect to the almost complex structure $J_m{}^n$, i.e. $f^{(1,0)}_{1|m}=\left(\Pi^+\right)_m{}^nf^{(1,0)}_{1|n}$.

{$\bullet$  Real three-form}
\eq{\label{dec3}F_{mnp}=f^{(2,1)}_{3|mnp}+3f^{(1,0)}_{3|[m}J_{np]}
+\tilde{f}^{(1,0)}_{3|s}\Omega^{s*}{}_{mnp}
+\mathrm{c.c.}~,}
where $f^{(2,1)}_{3|mnp}\sim \bf{20}$ is a complex traceless (2,1)-form, ${f}^{(1,0)}_{3|m}, \tilde{f}^{(1,0)}_{3|m}\sim \bf{4}$ are complex (1,0)-forms.
%
%

For the RR-forms $F_p$ with $p=5,7$ we expand the Hodge duals $\star_8F_p$ exactly as above:

{$\bullet$  Real five-form}
\eq{\label{dec5}(\star_8F_5)_{mnp}=f^{(2,1)}_{5|mnp}+3f^{(1,0)}_{5|[m}J_{np]}
+\tilde{f}^{(1,0)}_{5|s}\Omega^{s*}{}_{mnp}
+\mathrm{c.c.}~,}

{$\bullet$  Real seven-form}
\eq{\label{dec7}(\star_8F_7)_m=f^{(1,0)}_{7|m}+\mathrm{c.c.}~;}

For the NSNS three-form $H$ we decompose similarly:
\eq{\label{b19}
H=e^{2A}\mathrm{vol}_2\wedge h_1+h_3
~,}
where as before $\mathrm{vol}_2$ is the unwarped volume
element of $\mathbb{R}^{1,1}$; $h_1$, $h_3$ are real one-, three-forms
on $\mathcal{M}_8$ respectively. These further decompose to irreducible
$su(4)$-modules:
\eq{h_{1|m}=h_{1|m}^{(1,0)}+\mathrm{c.c.}~,}
with ${h}^{(1,0)}_{1|m}\sim \bf{4}$ a complex (1,0)-form, and
\eq{
h_{3|mnp}=
h^{(2,1)}_{3|mnp}+3h^{(1,0)}_{3|[m}J_{np]}
+\tilde{h}^{(1,0)}_{3|s}\Omega^{s*}{}_{mnp}
+\mathrm{c.c.}~,}
where $h^{(2,1)}_{3|mnp}\sim \bf{20}$ is a complex traceless (2,1)-form, ${h}^{(1,0)}_{3|m}, \tilde{h}^{(1,0)}_{3|m}\sim \bf{4}$ are complex (1,0)-forms.

\subsection{Useful formul\ae{}}\label{app2}

The following useful identities can be proved  by Fierzing \cite{tsim}:
\eq{\spl{
\frac{1}{4!\times 2^4}~&\Omega_{rstu}\Omega^{*rstu}=1\\
\frac{1}{6\times 2^4}~&\Omega_{irst}\Omega^{*mrst}
=(\Pi^+)_{i}{}^{m}\\
\frac{1}{4\times 2^4}~&\Omega_{ijrs}\Omega^{*mnrs}
=(\Pi^+)_{[i}{}^{m}(\Pi^+)_{j]}{}^{n}\\
\frac{1}{6\times 2^4}~&\Omega_{ijkr}\Omega^{*mnpr}
=(\Pi^+)_{[i}{}^{m}(\Pi^+)_{j}{}^{n}(\Pi^+)_{k]}{}^{p}\\
\frac{1}{4!\times 2^4}~&\Omega_{ijkl}\Omega^{*mnpq}
=(\Pi^+)_{[i}{}^{m}(\Pi^+)_{j}{}^{n}(\Pi^+)_{k}{}^{p}(\Pi^+)_{l]}{}^{q}~,
\label{bfive}
}}
Moreover, we have
\eq{\spl{
\widetilde{\eta^c}\eta=1; &~~~~~\we\eta=0\\
\widetilde{\eta^c}\g_{mn}\eta=iJ_{mn}; &~~~~~\we\g_{mn}\eta=0\\
 \widetilde{\eta^c}\g_{mnpq}\eta=-3J_{[mn}J_{pq]}; &~~~~~\we\g_{mnpq}\eta=\Omega_{mnpq}\\
\widetilde{\eta^c}\g_{mnpqrs}\eta=-15iJ_{[mn}J_{pq}J_{rs]}; &~~~~~\we\g_{mnpqrs}\eta=0\\
\widetilde{\eta^c}\g_{mnpqrstu}\eta=105J_{[mn}J_{pq}J_{rs}J_{tu]}; &~~~~~\we\g_{mnpqrstu}\eta=0 ~,
\label{usefids}
}}
where we have made use of the identities
\eq{\spl{\label{jvol}
\sqrt{g}\; \varepsilon_{mnpqrstu}J^{rs}J^{tu}&=24J_{[mn}J_{pq]}\\
\sqrt{g} \;\varepsilon_{mnpqrstu}J^{tu}&=30J_{[mn}J_{pq}J_{rs]}\\
\sqrt{g} \;\varepsilon_{mnpqrstu}&=105J_{[mn}J_{pq}J_{rs}J_{tu]}
~.
}}
Note that the bilinears
$\we\g_{(p)}\eta$, ~$\widetilde{\eta^c}\g_{(p)}\eta$, vanish for $p$ odd.
The last line of equation (\ref{bfive}) together with the last line of the
equation above imply
\eq{\label{omvol}
\Omega_{[ijkl}\Omega^*_{mnpq]}=\frac{8}{35}\sqrt{g}\; \varepsilon_{ijklmnpq}~.
}
%
%
%
Finally, the following relations are useful in the analysis of the Killing spinor equations:
\eq{\spl{
\g_m\eta&=(\Pi^+)_{m}{}^{n}\g_n\eta\\
\g_{mn}\eta&=iJ_{mn}\eta -\frac{1}{8}\Omega_{mnpq}\g^{pq}\eta^c   \\
\g_{mnp}\eta&=3iJ_{[mn}\g_{p]}\eta
-\frac{1}{2}\Omega_{mnpq}\g^q\eta^c\\
\g_{mnpq}\eta&=-3J_{[mn}J_{pq]}\eta -\frac{3i}{4}J_{[mn}\Omega_{pq]ij}\g^{ij}\eta^c
+\Omega_{mnpq}\eta^c
~.
\label{fierzsu}
}}
The action of $\gamma_{m_1\dots m_p}$, $p\geq 5$, on $\eta$ can be related to
the above formul{\ae}, using
the Hodge properties of gamma matrices given in appendix \ref{app1}.

\subsection*{Formul\ae{} needed for the dilatino equations}
In order to solve the dilatino equations, we make use of the following:
\eq{\spl{
\underline{h_1} \eta &= h_{1|m}^{(0,1)} \g^m \eta \\
\underline{h_3} \eta &= 3i h_{3|m}^{(0,1)} \g^m \eta
+8 \tilde{h}_{3|m}^{(1,0)} \g_m \eta^c
}}
and note that similar identities hold for the RR fluxes $F_1$, $\star_8 F_7$ and $F_3$, $\star_8  F_5$. \\
\\

\subsection*{Formul\ae{} needed for the internal gravitino equations}
For the gravitino equations with $M=m$, we require
\eq{
\underline{h_{3|m}}  \eta = 3i ( h^{(1,0)}_{3|m} + h^{(0,1)}_{3|m}) \eta -
\left(\frac{i}{8}h^{(0,1)}_{3|n}\O^{\phantom{m}nrs}_m + \frac{1}{16} h^{(1,2)}_{3|mpq} \O^{pqrs}\right)\g_{rs} \eta^c
-\frac{1}{2}\tilde{h}^{(1,0)}_{3|n}\O^{*\phantom{m}npq}_m\g_{pq} \eta
}
and
\eq{\spl{
  \underline{F_1} \g_m \eta &= 2 f_{1|m}^{(1,0)}  \eta + \frac{1}{8}f_{1|n}^{(0,1)} \O_m^{\phantom{m}npq}\g_{pq} \eta^c \\
  \underline{F_3} \g_m \eta &=   6i f_{3|m}^{(1,0)}\eta - 16 \tilde{f}_{3|m}^{(1,0)} \eta^c
                            + \left(
\frac{1}{8}i f_{3|n}^{(0,1)}  \O_m^{\phantom{m}nrs}
- \frac{1}{8} f_{3|mpq}^{(1,2)}  \O^{pqrs}
 \right)\g_{rs} \eta^c\;.
}}
Similar identities also hold for RR forms $\star_8 F_p$ with $p \geq 5$.

\section{\label{app:calibrations}Generalized geometry and generalized calibrations}

We here briefly introduce some relevant concepts from generalized complex geometry \cite{hitch,gual} and generalized calibrations that are used in the main text. The material in this section is not new: it is included here solely for the purpose of establishing conventions and making the paper self-contained. We refer to e.g. the review \cite{koer} for detailed explanations and references; for more details on the $\d^\mathcal{I}$ operator the reader may also consult \cite{caval}.

\subsection{\label{app:gengeom}Generalized geometry}

Generalized complex geometry is an extension of both complex and symplectic geometry interpolating, in a certain sense,  between these two special cases. Consider an even-dimensional
manifold $\cm_{2k}$. One can equip the sum of tangent and cotangent bundles $T\oplus T^*$ with a metric $\mathcal{H}$ of maximally indefinite signature  (the pairing between vectors and forms),
\eq{\mathcal{H}= \frac{1}{2}\left( \begin{array}{cc}
0 & \mathbbm{1} \\
\mathbbm{1} & 0 \end{array} \right) ~,}
reducing the structure group to $O(2k,2k)$. Imposing in addition the existence of an almost complex structure $\mathcal{I}$ on $T\oplus T^*$ associated with the metric $\mathcal{H}$ (i.e. such that $\mathcal{H}$ is hermitian with respect to $\mathcal{I}$:  $\mathcal{I}^T\cdot \mathcal{H}\cdot\mathcal{I}=\mathcal{H}$), further reduces the structure group to $U(k,k)$.

A pair $\mathcal{I}_{1,2}$ of {\it compatible} almost complex structures on $T\oplus T^*$ (i.e. such that they commute $[\mathcal{I}_{1},\mathcal{I}_{2}]=0$ and they give rise to a positive definite metric $\mathcal{G}=-\mathcal{I}_{1}\cdot\mathcal{I}_{2}$) further reduces the structure group to $U(k)\times U(k)$. The metric $\mathcal{G}$ on $T\oplus T^*$ associated with the pair $\mathcal{I}_{1,2}$ can be seen to give rise to both a positive definite metric $g$ and a $B$-field on $T$, via:
\eq{\mathcal{G}= \left( \begin{array}{cc}
\mathbbm{1} & 0 \\
B & \mathbbm{1} \end{array} \right)\cdot
 \left( \begin{array}{cc}
0 & g^{-1} \\
g & 0 \end{array} \right)\cdot
 \left( \begin{array}{cc}
\mathbbm{1} & 0 \\
-B & \mathbbm{1} \end{array} \right)~.}
Just as there is a correspondence between almost complex structures on $T$ and line bundles of pure Weyl spinors of $Cl(2k)$,\footnote{Recall that
pure Weyl spinors may be defined as the spinors which are annihilated by precisely those gamma matrices that are holomorphic (or antiholomorphic, depending on the convention) with respect to an almost complex structure.} there is a correspondence
\eq{\mathcal{I}\longleftrightarrow\Psi_{\mathcal{I}}~,\label{corr}}
between almost complex structures $\mathcal{I}$ on $T\oplus T^*$ and line bundles of pure spinors $\Psi_{\mathcal{I}}$ of $Cl(2k,2k)$.
More precisely, the $+i$ eigenbundle of $\mathcal{I}$ is isomorphic to the
space of generalized gamma matrices annihilating $\Psi_{\mathcal{I}}$.
Demanding that the line bundle of pure spinors of $Cl(2k,2k)$ has a global section, reduces the structure group of $T\oplus T^*$ from $U(k,k)$ (which was accomplished by the existence of a generalized almost complex structure) to $SU(k,k)$.

There is a natural action of $T\oplus T^{*}$ on the bundle $\Lambda^\bullet T^*$
of differential forms on $\cm_{2k}$, whereby every vector acts by contraction and every one-form by exterior multiplication. It can easily be seen that this action obeys the Clifford algebra $Cl(2k,2k)$ associated with the maximally indefinite metric $\mathcal{H}$ on $T\oplus T^*$. It follows that there is an isomorphism $Cl(2k,2k)\cong {\rm End}(\Lambda^\bullet T^*)$, which means that spinors on $T\oplus T^{*}$ can be identified with {\it polyforms} ({\it i.e.} sums of forms of different degrees) in $\Lambda^\bullet T^*$.

On the other hand, there is a correspondence between polyforms of $\Lambda^\bullet T^*$
and {\it bispinors} on $T$. Given a choice of the volume form, this correspondence is a canonical isomorphism, and is explicitly realized by the {\it Clifford map}: for two $Cl(2k)$ spinors $\psi_\a$, $\chi_\b$, one has
\eq{\label{cliffmap}
\psi_\alpha\otimes\widetilde{\chi}_\beta=\frac{1}{2^k}\sum_{p=0}^{2k}\frac{1}{p!}
(\widetilde{\chi}\gamma_{m_p\dots m_1}\psi)~\!\gamma^{m_1\dots m_p}_{\alpha\beta}
\longleftrightarrow
\frac{1}{2^k}\sum_{p=0}^{2k}\frac{1}{p!}
(\widetilde{\chi}\gamma_{m_p\dots m_1}\psi)~\!e^{m_1}\wedge\dots\wedge e^{m_p}
~,}
where the first equality is the Fierz identity.

It follows from the above discussion that the condition of compatibility
of a pair of generalized almost complex structures should be expressible as
a condition of compatibility on a pair of (line bundles of) pure spinors of $Cl(2k,2k)$ -- which, as already mentioned,  can
alternatively be thought of as either bispinors of $Cl(2k)$ or,
through \eqref{cliffmap}, as polyforms. Indeed, up to a factor and up to a $B$ transform ($\Psi_i\rightarrow e^B\wedge\Psi_i$, $i=1,2$), the most general
pair $\Psi_{1,2}$ of compatible pure spinors  of $Cl(2k,2k)$ is of the form:
\eq{\spl{\label{purespinors}
\Psi_1&= - \frac{2^k}{|a|^2}\eta_1\otimes\widetilde{\eta^c_2}\\
\Psi_2&= - \frac{2^k}{|a|^2}\eta_1\otimes\widetilde{\eta}_2
~,}}
where $\eta_{1,2}$ are pure spinors\footnote{\label{f1}Note that
for $k\leq3$, Weyl spinors of $Cl(2k)$ are automatically pure.
For the case $k=4$ one has to impose in addition
one complex condition, see section \ref{sec3}.} of $Cl(2k)$.
In the normalization above we have taken into account that the background admits calibrated branes, in which case $\eta_{1,2}$ have equal norm: $|a|^2=\widetilde{\eta}_1\eta_1^c=\widetilde{\eta}_2\eta_2^c$.

Provided the pair of pure spinors above is globally defined and nowhere vanishing (in other words: if the corresponding line bundles of pure spinors have nowhere-vanishing global sections), the structure group of $T\oplus T^*$ is further reduced from $U(k)\times U(k)$ (which was accomplished by the existence of a pair of compatible generalized almost complex structures) to $SU(k)\times SU(k)$.

The correspondence between generalized almost complex structures
and pure spinors allows one to express the condition of integrability of a
generalized almost complex structure as a certain first-order differential equation for the associated pure spinor, which may then also be called integrable. A manifold $\cm_{2k}$ is called {\it generalized complex} if it admits an integrable pure spinor.
A {\it generalized Calabi-Yau (GCY)} is a special case of a generalized complex manifold. It is defined as a manifold $\cm_{2k}$ on which a pure spinor $\Psi$ exists, obeying the differential condition\footnote{This is also sometimes called the `twisted' Calabi-Yau condition; the pure spinor
$\Psi$ is thought of as a polyform in $\Lambda^\bullet T^*$ via the
Clifford map (\ref{cliffmap}).}
\eq{\label{gcy}
\mathrm{d}_H\Psi=0~,}
where $\mathrm{d}_H\equiv\d+H\wedge~\!$ and $H=\d B$ is the field strength of the $B$ field.

Just as one can construct the components of any ordinary spinor by acting with holomorphic (or antiholomorphic, depending on the convention) gamma matrices on the Clifford vacuum, one can construct any polyform by acting on a generalized pure spinor vacuum with generalized gamma matrices which are holomorphic with respect to the almost complex structure $\mathcal{I}$ associated with that pure spinor  $\Psi$, cf.~(\ref{corr}).
Using this Fock space construction it can then be seen that there is a natural decomposition of polyforms $\Psi$:
\eq{\label{polyformdec}\Psi=\sum_{q=-k}^k\Psi^{(q)}~,}
so that $\Psi^{(q)}$ has $+iq$ eigenvalue with respect to  $\mathcal{I}$.\footnote{Similarly, it can be seen that in the case of a generalized $SU(k)\times SU(k)$ structure the existence of a compatible pair of pure spinors $\Psi_{1,2}$ implies a double decomposition of polyforms:
\eq{\Psi^{(q_1)}=\Psi^{(q_1,|q_1|-k)}+
\Psi^{(q_1,|q_1|-k+2)}+\dots+\Psi^{(q_1,k-|q_1|)}
~,}
where now $\Psi^{(q_1,q_2)}$  has $+iq_1$ eigenvalue with respect to $\mathcal{I}_1$ and $+iq_2$ eigenvalue with respect to $\mathcal{I}_2$.}

Let us further assume that the generalized almost complex structure  is integrable. This is indeed the case for the  almost complex structure $\mathcal{I}_2$ associated with the pure spinor $e^{2 A- \Phi} \Psi_2$ for the supersymmetric backgrounds considered in the present paper, cf. the second line of \eqref{conjp} and the discussion preceding \eqref{gcy}. ($\mathcal{I}_2$ is also the almost complex structure associated with the pure spinor $\Psi_2$, as can be seen from the fact that both $\Psi_2$ and $f \Psi_2$ have the same annihilator space for any function $f$.)
It then follows that the twisted differential $\d_H$ maps $q$-polyforms to the space of $(q+1)\oplus(q-1)$-polyforms:
\eq{
\d_H (\Psi^{(q)} )=\left(\d_H\Psi\right)^{(q+1)}
+\left(\d_H\Psi\right)^{(q-1)}
~.}
We can thus define a twisted {\it generalized Dolbeault operator} $\partial_H^{\mathcal{I}_2}$ associated with the integrable almost complex structure $\mathcal{I}_2$ via
\eq{\label{dolbeault}\partial_H^{\mathcal{I}_2}\Psi\equiv\left(\d_H\Psi\right)^{(q+1)}~,
~~~~~\bar{\partial}_H^{\mathcal{I}_2}\Psi\equiv\left(\d_H\Psi\right)^{(q-1)}~.}
It is also straightforward to see that
the twisted differential $\d_H^{\mathcal{I}_2}\equiv i(\bar{\partial}_H^{\mathcal{I}_2}-\partial_H^{\mathcal{I}_2})$ that appears in
\eqref{missing} admits an alternative equivalent definition:
\eq{\label{c12}\d_H^{\mathcal{I}_2}\equiv[\d_H,\mathcal{I}_2\cdot]~,}
where the operator $\mathcal{I}_2\cdot$ is defined via
$\mathcal{I}_2\cdot\Psi^{(q)}=iq\Psi^{(q)}$.
In the context of type II supergravity on backgrounds of the form $\mathbb{R}^{1,3}\times\mcal_6$ the operator $\d_H^{\mathcal{I}_2}$ has been further studied in \cite{toma}.

In the case of backgrounds of strict $SU(4)$ structure considered here, we can make contact with the ordinary Dolbeault operator as follows:
As can be seen from the second line of \eqref{4.1},  $\Psi_2$ is annihilated by acting on the left or the right with ordinary gamma matrices which are holomorphic with respect to the complex structure on $\mcal_8$. Using the generalized almost complex structure/annihilator space correspondence (\ref{corr}), this corresponds to the action of holomorphic (with respect to $\mathcal{I}_2$) generalized gamma matrices. This further corresponds, using the Clifford map, to contraction with an antiholomorphic vector or wedging with a holomorphic one-form. Hence the action of a holomorphic generalized gamma matrix  transforms an ordinary $(p,q)$-form (where now this refers to the ordinary Hodge decomposition) to a $(p+1,q)$-form and/or a $(p,q-1)$-form, i.e. it increases $(p-q)$ by one. On the other hand the action of a holomorphic generalized gamma matrix transforms a $\Psi^{(k)}$-polyform to a $\Psi^{(k+1)}$-polyform, cf. the discussion around (\ref{polyformdec}).
Similarly the action of an antiholomorphic generalized gamma matrix on an ordinary $(p,q)$-form dereases $(p-q)$ by one; on a $\Psi^{(k)}$-polyform it decreases $k$ by one.
Since the action of the ordinary twisted Dolbeault operator $\partial_H\equiv\partial+H^{(2,1)}$ on an ordinary $(p,q)$-form increases $(p-q)$ by one while the action of $\bar{\partial}_H\equiv\bar{\partial}+H^{(1,2)}$ decreases $(p-q)$ by one, we conclude from (\ref{dolbeault}) that for the strict $SU(4)$ structure considered here the generalized Dolbeault operator reduces to the ordinary twisted Dolbeault operator.

\subsection{\label{app:calib}Generalized calibrations}

The close connection between supersymmetry and  calibrations \cite{harvey} was noted some time ago \cite{cala,calb,calc,marino}. More recently, generalized calibrations in flux backgrounds were shown to have a natural interpretation
in terms of generalized geometry \cite{calmart,calkoer,koermart}. In this section we will briefly review the relevant results, referring the reader to \cite{koer} or the original literature for further details.

Consider the energy density $\mathcal{E}(\Sigma,\mathcal{F})$ of a static, magnetic ({\it i.e.} without electric worldvolume flux) D-brane in our setup, filling $q$ external spacetime dimensions and  wrapping a cycle $\Sigma$ in the internal space (for our purposes it will not be necessary to take higher-order corrections into consideration):
\eq{\label{endens}
\mathcal{E}(\Sigma,\mathcal{F})
=
e^{qA-\Phi}\sqrt{\det(g+\mathcal{F})}-\delta_{q,d}
\left(C^{\mathrm{el}}\wedge e^\mathcal{F}\right)_\Sigma
~,}
where $g$ is the induced worldvolume metric on $\Sigma$, $\mathcal{F}$ is the worldvolume flux: $d\mathcal{F}=H|_\Sigma$, and  $C^{\mathrm{el}}$ is the electric RR flux potential: $\mathrm{d}_HC^{\mathrm{el}}=F^{\mathrm{el}}$, cf. \eqref{fluxan}. Note that unless the brane fills all the external spacetime directions, the second term on the right hand side above vanishes. This property of the energy density follows from the form of the ansatz for the RR fields,  \eqref{fluxan}, which is such that it preserves the $d$-dimensional Poincar\'e invariance of the background.

A polyform $\omega$ (defined in the whole of the internal space) is a {\it generalized calibration form} if, for any cycle $\Sigma$, it satisfies the algebraic inequality:
\eq{\label{calalg}
\left(\omega\wedge e^{\mathcal{F}}\right)_\Sigma\leq \mathrm{d}\sigma e^{qA-\Phi}\sqrt{\det(g+\mathcal{F})}
~,}
where $\sigma$ collectively denotes the coordinates of $\Sigma$, together with the differential condition:\footnote{
Alternatively the calibration form is sometimes defined to obey:
\eq{
\left(\omega'\wedge e^{\mathcal{F}}\right)_\Sigma\leq \mathrm{d}\sigma\mathcal{E}(\Sigma,\mathcal{F})
~,\nn}
as well as the differential condition:
\eq{
\mathrm{d}_H\omega'=0
~.\nn
}
The two definitions are related by: $\omega'=\omega-\delta_{q,d}C^{\mathrm{el}}$; the one we adopt in the main text is more natural from the point of view of the calibrations/background supersymmetry correspondence.}
\eq{\label{caldiff}
\mathrm{d}_H\omega=\delta_{q,d}F^{\mathrm{el}}
~.}
A generalized submanifold $(\Sigma,\mathcal{F})$ is called {\it calibrated by $\omega$}, if it saturates the bound given in \eqref{calalg} above.

The upshot of the above discussion is that D-branes wrapping generalized calibrated submanifolds minimize their energy within their (generalized) homology class. Recall that $(\Sigma,\mathcal{F})$,  $(\Sigma',\mathcal{F}')$ are in the same generalized homology class if there is a cycle $\widetilde{\Sigma}$ such that $\partial\widetilde{\Sigma}=\Sigma'-\Sigma$
and there exists an extension of the worldvolume flux $\widetilde{\mathcal{F}}$ on $\widetilde{\Sigma}$ such that: $\widetilde{\mathcal{F}}|_\Sigma=\mathcal{F}$ and  $\widetilde{\mathcal{F}}|_{\Sigma'}=\mathcal{F}'$.
Then, if $(\Sigma,\mathcal{F})$ is calibrated by $\omega$ we have, using
Stokes theorem as well as eqs.~(\ref{endens}-\ref{caldiff}):
\eq{
\int_{\Sigma'} \mathrm{d}\sigma~\! \mathcal{E}(\Sigma',\mathcal{F}')\geq
\int\left(\omega-\delta_{q,d}C^{\mathrm{el}}\right)_{\Sigma'}\wedge e^{\mathcal{F}'}
=\int \left(\omega-\delta_{q,d}C^{\mathrm{el}}\right)_\Sigma\wedge e^\mathcal{F}
=\int_\Sigma \mathrm{d}\sigma ~\! \mathcal{E}(\Sigma,\mathcal{F})
~.}
In the special case of the backgrounds considered in the present paper the static, magnetic D-branes where shown in \cite{patalong} to be necessarily spacetime-filling (i.e. they wrap the external $\mathbb{R}^{1,1}$ part). The corresponding calibration form $\o$ is given by
\begin{align}
\o = e^{2 A - \phi} \Re( e^{i \varphi} \Psi_2 ) + e^{2 A - \phi} \Re \Psi_1
\end{align}
for some phase $\varphi$.

\end{document}